\documentclass[reprint, superscriptaddress, showkeys, amsmath, amssymb, aps, prl]{revtex4-1}
\usepackage{graphicx}
\usepackage{dcolumn}
\usepackage{bm}

\makeatletter

\newcommand{\Rmnum}[1]{\expandafter\@slowromancap\romannumeral #1@}
\makeatother

\begin{document}

\title{Gate-tunable strong-weak localization transition in few-layer black phosphorus}

\author{Gen Long}

\author{Shuigang Xu}
\author{Xiangbin Cai}
\author{Zefei Wu}
\author{Tianyi Han}
\author{Jiangxiazi Lin}
\author{Yuanwei Wang}
\author{Liheng An}
\author{Yuan Cai}
\affiliation{Department of Physics and Center for Quantum Materials, the Hong Kong University of Science and Technology, Hong Kong, China}

\author{Xinran Wang}
\affiliation{National Laboratory of Solid State Microstructures, School of Electronic Science and Engineering, and Collaborative Innovation Center of Advanced Microstructures, Nanjing University, Nanjing 210093, China}

\author{Ning Wang}
\email[Correspondence to: ]{phwang@ust.hk}
\affiliation{Department of Physics and Center for Quantum Materials, the Hong Kong University of Science and Technology, Hong Kong, China}

\date{\today}

\begin{abstract}

Atomically thin black phosphorus (BP) field-effect transistors show strong-weak localization transition which is tunable through gate voltages. Hopping transports through charge impurity induced localized states are measured at low-carrier density regime. Variable-range hopping model is applied to simulate the charge carrier scattering behavior. In the high-carrier concentration regime, a negative magnetoresistance signals the weak localization effect. The extracted phase coherence length is power-law temperature dependent ($\sim T^{-0.48\pm0.03}$) and demonstrates electron-electron interactions in few-layer BP. The competition between the Strong localization length and phase coherence length is proposed and discussed based on the observed gate tunable strong-weak localization transition in few-layer BP.

\end{abstract}


\maketitle

Atomically thin two-dimensional (2D) materials, such as graphene and transition metal dichalcogenides have opened new avenues for exploring physical property anomalies and potential applications in nanoelectronics and optoelectronics \cite{novoselov2005two, neto2009electronic, geim2009graphene, balandin2008superior, mattheiss1973band, wilson1969transition, ayari2007realization, yu2016realization, cui2015high, qiu2013hopping, qiu2012electrical, mak2010atomically, long2016charge, li2014black, xia2014rediscovering, qiao2014high}. Single- or few-layer black phosphorus (BP) \cite{li2014black, xia2014rediscovering, qiao2014high, castellanos2014isolation, takao1981electronic, ling2015renaissance} is another promising 2D material characterized by high mobility at room temperature, a tunable direct band gap, and high in-plane anisotropy properties for both the fundamental study of thermal, optical, and optoelectronics properties and technological applications \cite{balandin2008superior, yuan2015polarization, buscema2014photovoltaic, wang2015highly}. Recently, interesting Shubnikov de Haas oscillations and quantum Hall effects have been demonstrated in boron nitride (BN)-encapsulated few-layer BP high-mobility field-effect transistors (FETs) at cryogenic temperatures \cite{gillgren2014gate, long2016type, li2015quantum, tayari2015two, li2016quantum, long2016achieving}. However, achieving ultrahigh mobility BP FETs at low carrier density remains a challenge in the further study of a number of intriguing physics properties, such as many-body phenomena or fraction quantum Hall effects.

\par We discuss here an experimental study of the charge carrier scattering behaviors in few-layer BP at various charge carrier densities, temperatures, and magnetic fields. We found that at low-carrier concentration regime ($\sim10^{11} cm^{−2}$), strong localization is on account of low-carrier mobility of a few $cm^2V^{-1}s^{-1}$ and the observed localization effects can be fitted by the variable-range hopping (VRH) model for low temperature region \cite{abou1973selfconsistent}. Meanwhile, at high-carrier concentration ($\sim10^{12} cm^{−2}$), weak localization occurs and a high carrier mobility of up to $10^3 cm^2V^{-1}s^{-1}$ is achieved. We discuss the mechanisms for gate-tunable strong–weak localization transition by extracting the phase coherence lengths and inelastic scattering time in high-carrier concentrations \cite{altshuler1980magnetoresistance}.

\par Our few-layer BP FETs are fabricated using exfoliated BP flakes. The quality of BP flakes degrades in atmospheric conditions due primarily to oxidation. To avoid BP quality degradation, few-layer BP samples are encapsulated between two hexagonal BN (h-BN) sheets to form h-BN/BP/h-BN encapsulation structures \cite{gillgren2014gate, long2016type, li2016quantum, long2016achieving, chen2015high}. The exfoliation and encapsulation processes are performed in a glove box filled with nitrogen to protect the exfoliated BP flakes (more details regarding the device-fabrication process are available in the Method section).

\par
\begin{figure*}[!htb]
\includegraphics[scale=1]{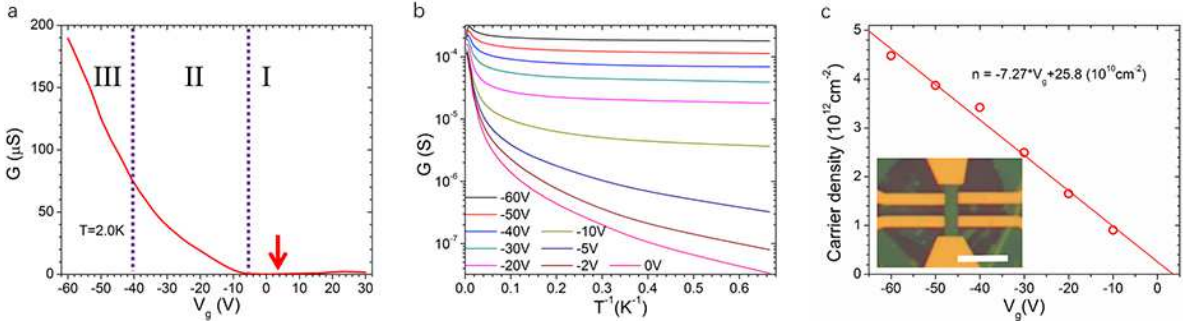}
\caption{\textbf{Carrier concentrations and temperatures dependence of channel performances.}  (a) Four-terminal transfer curve at T=1.8K. The red arrow indicates the charge neutral point (threshold gate voltage). The two vertical dashed lines divide the transfer curve into three different regions. (b) Arrhenius plot of channel conductance. (c) Carrier density obtained from Hall effect as a function of gate voltages. The inset shows an optical image of a typical device. Scale bar: 10$\mu m$}
\label{Fig.1} 
\end{figure*}

 Figure 1a shows the variation of BP channel conductance for different gate voltages at 1.8 K. The red arrow indicates the charge neutrality point (CNP) obtained from the linear fitting of the charge carrier versus the gate voltage (Fig. \ref{Fig.1}c). The charge carrier concentrations shown in Fig. \ref{Fig.1}c are determined by Hall effect measurement. The vertical dashed lines in Fig. \ref{Fig.1}a divide the transport curve into three distinguishable regions. In region \Rmnum{1}(adjacent to CNP), the carrier mobility ranges within a few $cm^2V^{-1}s^{-1}$. A high mobility of approximately $2000 cm^2 V^{−1} s^{−1}$ is achieved in region \Rmnum{3} (high-carrier concentration), indicating different charge carrier scattering behaviors at different carrier concentrations in our BP FETs. The transition between the two different scattering behaviors occurs in region \Rmnum{2}.

\par To further explore the different charge carriers scattering behaviors, we measured the temperature T dependences of channel conductance G for different gate voltages as shown in Fig. \ref{Fig.1}b. For low-carrier concentrations ($V_g > -5V$), decreasing the temperature strongly influences channel conductance when the temperature surpasses 100 K. On the other hand, the impact of temperature on channel conductance is much smaller for temperatures lower than 40 K. These temperature-dependence features of channel conductance indicate strong localizations induced by charge impurities for low-carrier concentrations in our devices. For high-carrier concentrations ($V_g <-40V$), the channel conductance shows a weak dependence on temperature, and the magnetic resistance of our devices shows typically weak localization features. To further confirm the claimed gate-tunable strong–weak localization transition in few-layer BP, the charge-carrier scattering behaviors are simulated by VRH, and Hikami–Larkin–Nagaoka (HLN) models at low- and high-carrier concentrations.

\section{Strong localization at low-carrier concentrations ($\sim10^{11} cm^{−2}$)}
\begin{figure}[!htb]
\includegraphics[width=8cm]{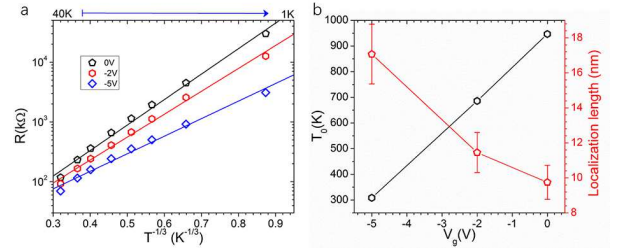}
\caption{\textbf{Variable-range hopping at low temperature.}  (a) Channel resistances as functions of $T^{-1/3}$ for different gate voltages (carrier concentrations). The solid lines show the fitting results of variable-range hopping model. (b) $T_0$ and localization lengths obtained from the fitting results.}
\label{Fig.2} 
\end{figure}

Figure \ref{Fig.2}a shows the channel resistance plotted as a function of $T^{−1/3}$ at low-carrier concentrations at temperatures ranging from 1.8 K to 40 K. The linear dependence of $log R$ vs. $T^{−1/3}$ perfectly fits the 2D VRH model:
\[ R\varpropto exp(\frac{T_0}{T})^{\frac{1}{2+1}}  \]
where $T_0$ is the characteristic temperature \cite{hill1976variable}. The gate voltage dependence of $T_0$ (Fig. \ref{Fig.2}b) agrees with the VRH model and serves as the signature of hopping transport via localized states \cite{qiu2013hopping, han2010electron}. The localization length $\xi_{VRH}$ can be extracted from $\xi_{VRH}= \sqrt{\frac{13.8}{k_B \rho(E) T_0}}$, where $T_0$ is the characteristic temperature obtained from the VRH model and $\rho(E)$ is the density of state (DOS) at the Fermi level \cite{li2011electron}. The DOS of the two dimensional hole gas can be expressed as $\rho(E)=\frac{m^\ast}{\pi \hbar^2}$ , where $m^\ast=0.26 m_0$ is the effective mass of holes. The calculated gate voltage dependence of $\xi_{VRH}$ is shown in Fig. \ref{Fig.2}b. $\xi_{VRH}$ increases as the gate voltage decreases (increasing the carrier concentration), a pattern that is widely observed in two dimensional electron systems \cite{han2010electron, yu2001variable}. The increasing trend of  $\xi_{VRH}$ can be explained by the two dimensional hydrogen atom model. The high energy states are occupied when the density of the localized charge carriers increases, resulting in two dimensional hydrogen atoms with a large radius (localization length). From another perspective, the increasing localization length may also be explained by the screening effect, as the localized charge carriers screen the electronic field of the localization center and thus contribute to a weak localization strength, i.e., large localization length \cite{qiu2013hopping, van1997screening}.

\par As temperature increases, the resistance deviates from the 2D VRH model and fits to the Arrhenius behavior $R\propto exp(\frac{T_1}{T}$ as shown in Fig. S1a (Supplementary materials), where $T_1$ is the characteristic temperature which reflects excitation energy. Normally, the linear dependence of $log R$ on $T^{-1}$ signals the crossover from VRH to nearest neighbor Hopping model. However in our device, the excitation energy extracted from the fitting results is quite small (Fig. S1b, Supplementary materials) which is lower than the physical temperature. The scattering process is rather dominated by the high-temperature incoherence diffusive transport than nearest neighbor hopping \cite{gershenson2000hot}. 

\section{Weak localization at high-carrier concentrations ($\sim10^{12} cm^{−2}$)}
\begin{figure}[!htb]
\includegraphics[width=8cm]{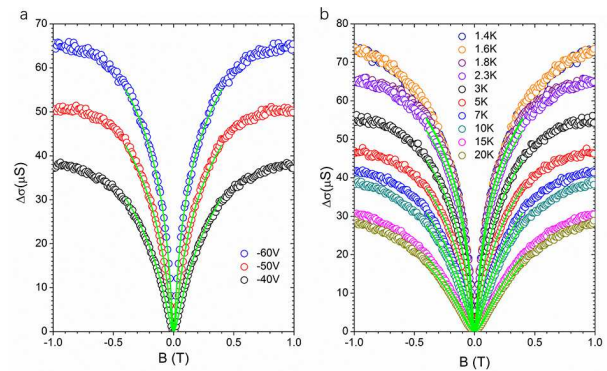}
\caption{\textbf{Weak localization at high carrier concentrations.}  (a) and (b) Normalized conductivity $\delta \sigma$ as functions of magnetic fields for different gate voltages at $T=1.8K$, and for different temperatures with a fixed gate voltage of $V_g=-60V$, respectively. The green solid lines show the fitting results of Hikami-Larkin-Nagaoka (HLN) model.}
\label{Fig.3} 
\end{figure}

As carrier concentration further increases, the dependence of the channel resistance on temperature becomes increasingly weak (Fig. \ref{Fig.1}b). The VRH model fails to describe the variation of the channel resistance after entering region \Rmnum{2} (Fig. \ref{Fig.1}a). To explore the transport behavior of charge carriers in the high-carrier concentration regime ($\sim10^{12} cm^{−2}$), we measured the magneto-conductance of our samples under different gate voltages and temperatures (Figs. \ref{Fig.3}a and b). The applied perpendicular magnetic field leads to a positive magneto-conductance (negative magneto-resistance), which is consistent with the features of weak localization \cite{van1985observation, du2016weak, shi2016weak}. The weak localization is a quantum correction to the conductance of a diffusive system originating from the phase interference of charge carrier wave functions. For a 2D electron system with a zero Berry phase $\phi$, the wave functions of backscattered charge carriers would have constructive interference, which increases of the probability of backscattering.    Thus, weak localization generally induces a negative quantum correction to the channel conductance. The applied perpendicular magnetic field induces extra changes to $\phi$, which break the constructive interference of backscattering and consequently lead to the positive magneto-conductance. The weak localization is strongly carrier-concentration dependent. In the present case, the magneto-conductance increases with the increasing carrier concentration (Fig. \ref{Fig.3}a). When the temperature is lower than 2 K, the weak localization is weakly temperature dependent. The magneto-conductance will decrease with increasing temperature when $T > 2 K$. To further explore the carrier concentration and temperature dependences of weak localization, we simulate the measured magneto-conductance with the HLN model:
\[\delta \sigma = \sigma(B)-\sigma(B=0) = \frac{e^2}{\pi h} [\Psi(\frac{1}{2}+\frac{B_\phi}{B})-ln(\frac{B_\phi}{B})] \]
where $\sigma$ is the device conductivity; $h$ is the Planck constant; and $e$ is the elemental charge. $\Psi$ refers to the digamma function. $B_\phi=\frac{\hbar}{4eL_\phi^2}$ is the phase coherence magnetic field, and $L_\phi$ is the phase coherence length \cite{altshuler1982effects, bergmann1984weak, hikami1980spin}. The green solid lines in Figs. \ref{Fig.3}a and b represent the carrier concentration and temperature fitting results, respectively, of the HLN model. The HLN model reproduces the experimental features of magneto-conductance between −0.4 and 0.4 T, especially the deep dips at zero magnetic field.

\par
\begin{figure*}[!htb]
\includegraphics[scale=1]{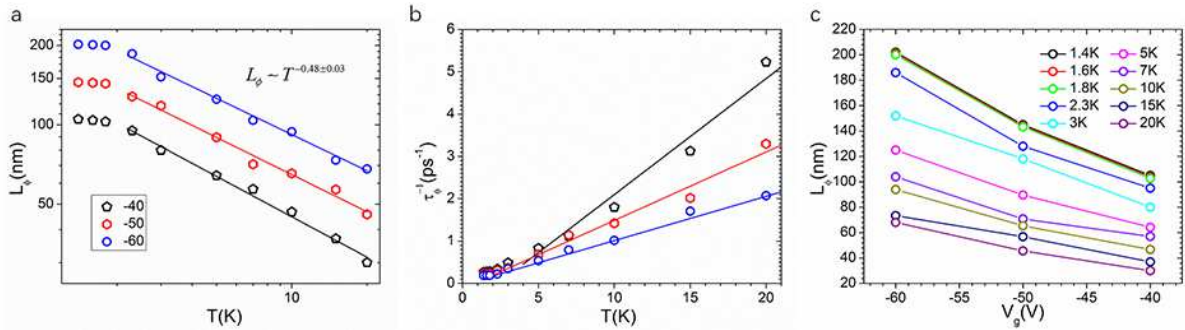}
\caption{\textbf{Phase coherence lengths $L_\phi$ and inelastic scattering time $\tau_\phi$ obtained from Hikami-Larkin-Nagaoka (HLN) model.}  (a) $L_\phi$ depends on temperatures at different gate voltages. (b) $\tau_\phi^{-1}$  linearly depends on temperatures for different gate voltages. (c) $L_\phi$ depend on gate voltages at varying temperatures. }
\label{Fig.4} 
\end{figure*}

 Figure \ref{Fig.4}a displays the temperature dependence of  $L_\phi$ for various gate voltages. A maximum $L_\phi$ of $202 nm$ is obtained at $T=1.4 K$ and $V_g=-60V$. For temperatures lower than 2 K, $L_\phi$ exhibits weak dependence on temperature. We fit  $L_\phi$ with a power–law formula $L_\phi\propto T^{-\beta}$ from 2.3 K to 40 K. A universal $\beta=0.48\pm0.03$ for different gate voltages is obtained from the fitting results. Specifically, the temperature dependence of inelastic scattering time $\tau_\phi=\frac{L_\phi^2}{D}\propto T^{-\alpha}$ distinguishes different scattering mechanisms, where $\alpha=2\beta$ and $D=\frac{\sigma \pi \hbar^2}{m^\ast e^2}$ is the diffusion constant. $\tau_\phi$ is determined to be 5.3 ps at T=1.4 K and $V_g=-60V$. The linear dependence of $\tau_\phi^-1$ on $T$ as shown in Fig. \ref{Fig.4}b demonstrates that $\alpha = 1$, which is consistent with the obtained $\beta$ value. The inelastic electronic interactions with small momentum transfer can be described by the Altshuler Aronov Khmelnitsky theory $\tau_\phi=\frac{\hbar}{k_B T} \frac{h\sigma/e^2}{ln(h\sigma/e^2}$, and the linear dependence of $\tau_\phi^{-1}$ on $T$ can be expected from the theory considering channel conductance as independent on temperature. The observed deviation of $\beta$ from 0.5 can be explained by the temperature dependence of $\sigma$ \cite{altshuler1982effects}. The gate voltage (or carrier concentration) dependence of $L_\phi$ is depicted in Fig. \ref{Fig.4}c, and $L_\phi$ decreases with decreasing carrier concentration. This decrease of  $L_\phi$ can be ascribed to the carrier-concentration dependence of $\sigma$.

The observed gate tunable strong–weak localization transition can be ascribed to the competition between  $L_\phi$ and $\xi_{VRH}$. At low-carrier concentrations, $\xi_{VRH}< L_\phi$ leads to a suppression of the interference of electronic wave functions, and strong localization dominates the weak localization caused by electron–electron interactions \cite{moser2010magnetotransport, matis2012giant, gershenson2000hot}. As carrier concentrations increase,  $\xi_{VRH}$ increases and reaches $\xi_{VRH}> L_\phi$ at high-carrier concentrations. In this case, the electronic wave function interference plays the leading role in the charge-carrier scattering processes, and strong localization is suppressed \cite{moser2010magnetotransport, matis2012giant, gershenson2000hot}. While, in region \Rmnum{2} (Fig. \ref{Fig.1}a), both strong and weak localizations influence the transport properties of the BP samples.

This study probed distinguishable charge-carrier transport behaviors for varying carrier concentrations in few-layer BP. Strong and weak localization models are proposed to simulate charge-carrier scattering behavior for low- and high-carrier concentrations, respectively. The strong localization is confirmed by the observation of VRH at low temperatures. The weak localization model is confirmed by the magneto-transport features. A competition behavior between strong localization length and phase coherence length is proposed to occur on account of the gate-tunable strong–weak localization transition. 

\paragraph{}  \textbf{Method}
\newline Few-layer BP and h-BN thin films are exfoliated from the bulk crystals on heavily doped silicon substrates covered with 300 nm thick $SiO_2$. Another h-BN flake is simultaneously prepared on the poly(methyl methacrylate) (PMMA) thin film. The h-BN flake on the PMMA film is used to pick up the BP flakes, and the formed h-BN/BP structure is then placed on the h-BN flakes on a silicon substrate. The h-BN/BP/h-BN heterostructure is then constructed. All these processes are performed in an inert gas environment to minimize the degradation of BP quality. This method avoids the direct contact between the PMMA film and BP flakes. Annealing at $300 ^{\circ}C$ in Ar flow for 10 h is applied to further stabilize the heterostructure. 
\\ Reaction-ion etching (RIE) (recite: 4 sccm $O_2$ +40 sccm $CHF_3$; RF power: 200 W) after the electron beam lithography (EBL) is used to define the Hall structures. A selective RIE etching is then applied to etch the top h-BN at the contact areas defined by the second EBL. The third EBL is then used to define the contact metal patterns followed by the electron beam evaporation to deposit the contact metals (Cr/Au=5/60 nm).
\\ Electrical measurements are performed with the lock-in technique in a cryogenic system under low pressure.

\paragraph{}  \textbf{Acknowledgements} 
\newline We acknowledge the helpful discussion with Prof. Sheng.
\newline Financial support from the Research Grants Council of Hong Kong (Project Nos. 16302215, HKU9/CRF/13G, 604112, and N\_HKUST613/12)and technical support of the Raith-HKUST Nanotechnology Laboratory for the electron-beam lithography facility at MCPF are hereby acknowledged.

\paragraph{} \textbf{Competing financial interests}
\newline The authors declare no competing financial interests.

\paragraph{} \textbf{Author contributions}
\newline G. Long and N. Wang conceived the project. G. Long fabricated the devices and performed cryogenic measurements with the help of S. Xu. G. Long，X. Wang and N. Wang analyzed the data and wrote the manuscript. Other authors provided technical assistance in the project.

\bibliography{localization_BP}

\end{document}